\documentclass[plb,reprint]{revtex4-1}
\usepackage{amssymb,amsfonts,amsmath,graphicx}

\newcommand\comment[1]{}

\begin{document}


\title{An Elementary end of the Periodic Table}



\author{Yang-Hui He}
\affiliation{Merton College, University of Oxford, UK,}
\affiliation{Department of Mathematics, City, University of London, UK,}
\affiliation{School of Physics, NanKai University, China}
\email[]{hey@maths.ox.ac.uk}

\author{Stavros Garoufalidis}
\affiliation{School of Mathematics, Georgia Institute of Technology, Atlanta, Georgia 30332-0160, USA,}
\affiliation{Max Planck Institute for Mathematics, Vivatsgasse 7, 53111 Bonn, Germany}
\email[]{stavros@math.gatech.edu}


\date{\today}

\begin{abstract}
Using the Planck scale as an absolute bound of half-life, we give a quick estimate, in the manner of Feynman's fine-structure method, of the highest possible atomic number.
We find, upon simple extrapolation, that element 168  would constitute the end of the Periodic Table and its isotope with atomic weight 411, being the most stable.
These are remarkably close to current best estimates obtained from sophisticated and much more involved Hartree-Fock calculations.
\end{abstract}


\maketitle


\section{Introduction and Summary}
Whether there exists the heaviest element, i.e., whether there exists the last element with the highest atomic number $Z_max$ on the Periodic Table, is clearly a fundamental question of great importance. There have been numerous theoretical estimations on possible upper bounds 
(q.v.~e.g., \cite{py,em,sch}) as well as experimental progress in the creation of increasingly higher atomic numbers (e.g., \cite{em,og,kib,GB}). 

Here, combining an extrapolation upon the maximal half-life of the known isotopes of transuranic elements together with one of the most fundamental constants of nature, $t_P$, the Planck time, we give a simple argument as to why the Period Table might end around atomic number 168, whose most stable isotope has atomic weight around 411. 

To our knowledge, this is the first estimate on the periodic table putting together the regime of nuclear theory with the theory of elementary particles and quantum gravity, which usually would never be considered in conjunction because of their vast difference in energy scale. It is therefore surprising that such reasonable numbers as $(168, 411)$ could be reached using such an elementary and quick estimate, and which is not too far at all from much more sophisticated methods on a current best upper bound \cite{py}. Given the fundamental nature of the Planck-scale, our estimate would constitute an {\em absolute} theoretical bound, in the sense that the law of physics would place such an estimate as an ultimate one.

Attempted theoretical upper bound to the atomic number has a distinguished history going back at least to the beginnings of nuclear physics. Meitner and Frisch suggested that Z=100 is the limit after their discovery of induced fission \cite{FM}. This limit, modelled on the nucleus being a charged droplet, was pushed up by the incorporation of microscopic effects by Strutinsky \cite{st}. Feynman’s back-of-the-envelope estimate was based on the speed $v \simeq Zc \alpha$ of the 1s orbital electron (where $c$ is the speed of light and $\alpha≃1/137.036$ is the fine structure constant) not exceeding $c$, and thus the limit $Z_max \simeq 137$ was proposed. This estimate assumed a point-like nucleus and a more accurate calculation \cite{sch} taking nuclear size into account pushed $Z_max$ to approximately 173, as supported by Hartree-Fock-Slater methods \cite{py,FGW}. 

Since the first postulate of ``Islands of Stability'' by Myers and Swiatecki \cite{MS} where certain combinations of atomic number $Z$ and the number of neutrons $N$ tend to give more stable isotopes, there has much exploration in the plot of stability regions in the $(N,Z)$-plane \cite{em,He}. The largest $Z$ to date is 118, which has recently been recognized and named as Oganesson (Og) \cite{cere}. This is particularly significant because Og completes the 7p orbital and the next element would occupy an entire new row in the Periodic Table.

\section{Methodology}
Whilst there have been bounds using the ratio of atomic to molecular mass \cite{kh} as well as relativistic estimates \cite{em,pers}, we shall take an entirely different method which combines available data and a quick estimate in the spirit of Feynman, to arrive at surprisingly reasonable numbers. 

First, we note that starting from Uranium ($Z=92$), all the elements (so-called transuranic) are unstable and decay radioactively. While the decay rates are different for different isotopes, the maximal half-lives $\tau$ (i.e., the longest-lived observed isotopes) are well-known \cite{em,kib,MG}. 
For reference, we present $\tau$ (in seconds) for the isotopes of $Z=92$ until the highest known number of 118, together with the atomic weight $W$ in Table \ref{t:tau}.

\begin{table}[t!!]
\begin{tabular}{ccc}
\begin{tabular}{|c|c|} \hline
${}^{238}$U$_{92}$  & $1.41 \times 10^{17}$ \\
${}^{237}$Np$_{93}$ &  $6.75 \times 10^{17}$ \\
${}^{244}$Pu$_{94}$  &  $2.52 \times 10^{15}$  \\
${}^{243}$Am$_{95}$ & $2.33 \times 10^{11}$ \\
${}^247$Cm$_{96}$ & $4.92 \times^{14}$ \\
${}^{247}$Bk$_{97}$ &  $4.35\times 10^{10}$ \\
${}^{251}$Cf$_{98}$ & $2.83 \times 10^{10}$ \\
${}^{252}$Es$_{99}$ &  $4.08 \times 10^7$ \\
${}^{257}$Fm$_{100}$ & $8.68\times 10^6$ \\
${}^{258}$Md$_{101}$ & $4.43 \times 10^6$ \\
${}^{259}$No$_{102}$ & $3.48 \times 10^3$ \\
${}^{266}$Lr$_{103}$ & $2.64 \times 10^4$ \\
${}^{267}$Rf$_{104}$ & $1.20 \times 10^4$ \\
${}^{268}$Db$_{105}$ & $6.96 \times 10^4$ \\
\hline
\end{tabular}
&
\qquad
&
\begin{tabular}{|c|c|} \hline
${}^{269}$Sg$_{106}$ & $1.86 \times 10^2$ \\
${}^{270}$Bh$_{107}$ & $6.00 \times 10^1$ \\
${}^{270}$Hs$_{108}$ & $1.00 \times 10^1$ \\
${}^{278}$Mt$_{109}$ & $7.60$ \\
${}^{281}$Ds$_{110}$ & $9.60$ \\
${}^{282}$Rg$_{111}$ & $1.00 \times 10^2$ \\
${}^{285}$Cn$_{112}$ & $2.90 \times 10^1$ \\
${}^{286}$Nh$_{113}$ & $9.50$ \\
${}^{289}$Fl$_{114}$ & $1.90$ \\
${}^{290}$Mc$_{115}$ & $6.50 \times 10^{-4}$ \\
${}^{293}$Lv$_{116}$ & $5.70 \times 10^{-5}$ \\
${}^{294}$Ts$_{117}$ & $5.10 \times 10^{-5}$ \\
${}^{294}$Og$_{118}$ & $6.90 \times 10^{-7}$ \\
\hline
\end{tabular}
\end{tabular}
\caption{{\sf 
The longest half-life of the known transuranic elements.}
\label{t:tau}}
\end{table}

\begin{figure}
\centering
\includegraphics[trim=0mm 0mm 0mm 0mm, clip, width=3in]{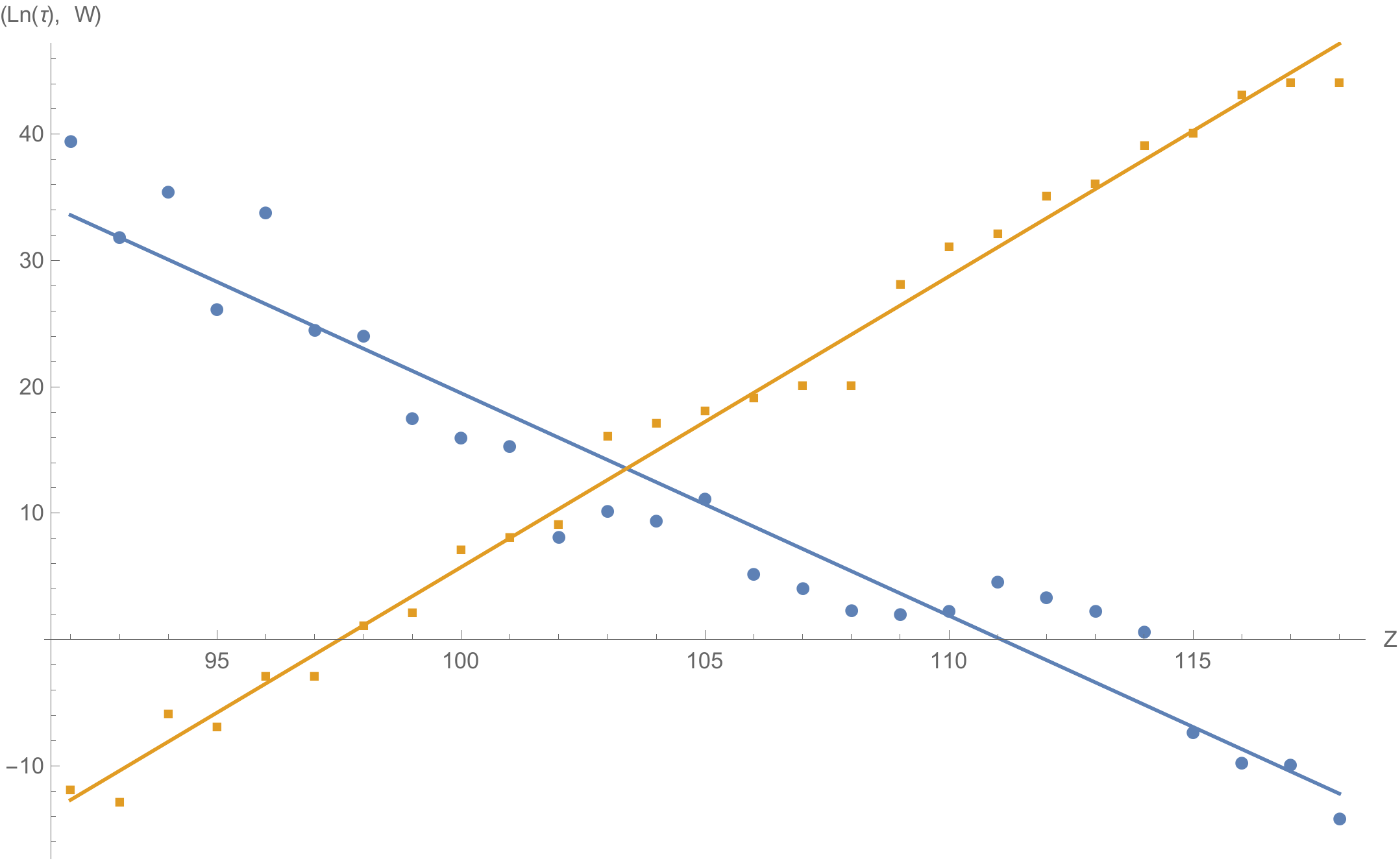}
\caption{{\sf A plot of the natural log of the maximal half-life $\tau$ and the atomic weight $W$ (subtracted by 250), against the atomic number $Z$.
\label{f:plot}
}}
\end{figure}

We plot $W-250$ (the shift is so that the ranges would be comparable) as well as the (natural) logarithm of $\tau$, both against $Z$ and obtain Figure \ref{f:plot}.

It is clear that both are linear to significant confidence. Specifically, we have that
 \begin{equation}
 \ln(\tau) \simeq 195.68 - 1.76 Z \ ,
 \end{equation}
  with p-value $8.37\times 10^{-16}$ and F - Statistic $322.69$, and that
  \begin{equation}
 W \simeq 25.43+2.30 Z \ ,
 \end{equation}
  with p-value $7.38 \times 10^{-27}$ and F - Statistic $2633.02$,
both suggestive of a good fit. 
We can therefore rather confidently extrapolate both maximal half-life and maximal atomic weight to beyond $Z=118$.

\section{Conclusions}
Now, there is a natural ``limit'' for time in the context of elementary particle physics which physical chemists and nuclear physicists normally do not consider and herein lies another novelty of our investigations. From the point of view of fundamental physics, Planck time is where the very notion of space-time needs dramatic modification, let alone the concepts of nuclei and atoms. This is the regime where a hypothetical unified theory of quantum gravity, of which the best candidate is string theory, exists. 

In seconds, Planck time is $t_P= \sqrt{\hbar G c^{-5}} \simeq 5.39 \times 10^{-44}$ where $G$ is Newton's universal gravitation constant and $\hbar$, the Planck constant.
Therefore, the natural logarithm of $t_P$ is a limit above which we cannot extrapolate our linear fit for $\ln(\tau)$. Upon substitution, we arrive at an elementary upper bound to the atomic number $Z_max \simeq 168$, which is surprisingly low and reasonable; indeed one might {\it ab initio} expect Planck limitations in elementary particle theory to give a much higher bound to atomic and nuclear quantities.
At this value, using our fit for atomic weight, we obtain $W=411$.

Recently, similar extrapolations have been performed in estimating the vast landscape coming from string theory compactifications \cite{He:2017aed,Constantin:2018xkj,Altman:2018zlc}.
In particular, a log-linear regression as performed in this letter was applied to known exact (minimally supersymmetric) Standard Models from heterotic string phenomenology \cite{Constantin:2018xkj} to arrive at an estimate of the number of vacua.

Finally, we remark  that usually within the regime of nuclear physics, ``existence'' of an isotope means a half-life around $10^{-14}$ seconds, roughly the scale of time taken for a nucleus to acquire its outer electrons \cite{bar}. Were we to use this at the upper-bound time, we would obtain a much smaller upper-bound of around $Z_{max} \simeq 129$. Of course, since we would now be very much working within the energy-scale of nuclear physics, such an extrapolation would be too naive and more sophisticated Hartree-Fock methods are needed. 

The key of this paper is to point out that a daring and back-of-the-envelop extrapolation to the lower bound of time-scale in elementary particle physics could produce so reasonable an answer in nuclear physics and give an absolute bound near (and in fact smaller) than the current best theoretical computations.

In conclusion, elementary considerations place element $X_{168}$ as the end of the Periodic Table, whose isotope 
\[
{}^{411}X_{168}
\] 
is expected to have the longest half-life.

\newpage

\begin{acknowledgments}
We are grateful to the organisers of the conference ``Geometry, Quantum Topology and Asymptotics 2018'' at L'institut confucius, l'universit\'e de Gen\`eve and at Sandbjerg Estate, University of Aarhus, which brought the convivial environment wherein the authors had many fruitful conversations.
YHH is indebted to the Science and Technology Facilities Council, UK, for grant ST/J00037X/1,
the Chinese Ministry of Education, for a Chang-Jiang Chair Professorship at NanKai University,
and the city of Tian-Jin for a Qian-Ren Award, as well as Merton College, Oxford for continued support. 
\end{acknowledgments}






\end{document}